\documentclass[apj]{emulateapj}
\slugcomment{Accepted for publication at ApJ Letters}
\usepackage{graphicx,amssymb,amsmath,times}

\begin{document}
\title{First bounds on the high-energy emission from isolated Wolf-Rayet binary systems}

\shorttitle{First bounds on the high-energy emission from isolated WR binary systems}
\shortauthors{Albert et al.}

%
\author{
E.~Aliu\altaffilmark{a},
H.~Anderhub\altaffilmark{b},
L.~A.~Antonelli\altaffilmark{c},
P.~Antoranz\altaffilmark{d},
M.~Backes\altaffilmark{e},
C.~Baixeras\altaffilmark{f},
J.~A.~Barrio\altaffilmark{d},
H.~Bartko\altaffilmark{g},
D.~Bastieri\altaffilmark{h},
J.~K.~Becker\altaffilmark{e}
% ...
}
\altaffiltext{a} {IFAE, Edifici Cn., Campus UAB, E-08193 Bellaterra, Spain}
\altaffiltext{b} {ETH Zurich, CH-8093 Switzerland}
\altaffiltext{c} {INAF National Institute for Astrophysics, I-00136 Rome, Italy}
\altaffiltext{d} {Universidad Complutense, E-28040 Madrid, Spain}
\altaffiltext{e} {Technische Universit\"at Dortmund, D-44221 Dortmund, Germany}
\altaffiltext{f} {Universitat Aut\`onoma de Barcelona, E-08193 Bellaterra, Spain}
\altaffiltext{g} {Max-Planck-Institut f\"ur Physik, D-80805 M\"unchen, Germany}
\altaffiltext{h} {Universit\`a di Padova and INFN, I-35131 Padova, Italy}
% ...

\begin{abstract}  
High-energy gamma-ray emission is theoretically expected to arise in
tight binary star systems (with high mass loss and high velocity
winds), although the evidence of this relationship has proven to be
elusive so far. Here we present the first bounds on this putative
emission from isolated Wolf-Rayet (WR) star binaries, WR\,147 and
WR\,146, obtained from observations with the MAGIC telescope.

\end{abstract}


\section{Introduction} 

WR stars represent an evolved stage of hot ($T_\textrm{eff} > 20000$\,K),
massive ($M_\textrm{ZAMS}>25 M_\odot$) stars, and display some of the
strongest sustained winds among galactic objects: their terminal
velocities may reach the range $v_\infty >1000-5000$\,km/s. Perhaps
with the exception of the short lived luminous blue variables (LBVs),
they have the highest known mass loss rate $\dot M \sim
10^{-4}...10^{-5}~M_\odot/$\,yr of any stellar type.  Thus, colliding
winds of massive star binary systems are considered as potential sites
of non-thermal high-energy photon production, via leptonic and/or
hadronic process after acceleration of primary particles in the
collision shock \citep[e.g.,][]{Eichler93}.  This possibility is
substantiated by the detection of (non-thermal) synchrotron
radio-emission from the expected colliding wind location in some
binaries (see below). Many models have been proposed to predict GeV to
TeV emission from these binaries, with different levels of detail
\citep[e.g., among recent works
see][]{Benaglia01,Benaglia03,Pittard06, Reimer06}. Conceptually, the
process would mimic the cases of LS\,5039 \citep{Aha05b}, PSR\,B1259-63
\citep{Aha06} or LS\,I\,+61\,303 \citep{Alb06,Alb08d,Alb08e}, particularly if
they result in pulsar-driven $\gamma$-ray binaries in which
$\gamma$-rays may arise from a shock region produced by the
interaction of the winds of the two components.\footnote{We recall
that these binaries have orbital modulated TeV emission, and are
formed by a massive star with a strong wind and a compact object,
which only in the case of PSR B 1250-63 is known to be a pulsar.}

In a recent paper, the High Energy Stereoscopic array (H.E.S.S.)
collaboration reported the discovery of very high energy (VHE)
$\gamma$-ray emission coincident with the young stellar cluster
Westerlund\,2 \citep{Aha07}.  The High Energy Gamma Ray Astronomy
(HEGRA) and the Major Atmospheric Gamma Imaging Cherenkov (MAGIC)
telescope detected the source TeV\,J2032+4130 and suggested a
possible connection with the Cygnus OB2 cluster
\citep{Aha02,Aha05a,Alb08a}.
Theoretically, the relationship between stellar associations and
high-energy emission has been put forward by many authors, since
early-type stellar associations have long been proposed as cosmic-ray
acceleration sites and also as providers of target material for
cosmic-ray interactions \citep[see
for instance the recent papers
by][]{
Parizot04,Torres04,Bednarek05,Eva06} and references therein.

In the case of Cygnus\,OB2, the VHE emission is supposed to occur at a
region displaced from the center of the association, where detailed
multiwavelength studies revealed an overdensity of hot OB stars,
although not WRs \citep{Butt03,Butt06}. In the case of Westerlund\,2,
the stellar cluster contains at least a dozen early-type O-stars, and
two remarkable WR stars, WR\,20a and WR\,20b. In particular, WR\,20a
was recently established to be a binary \citep{Rauw04,Bonanos04}.
Based on the orbital period, the minimum masses were found to be
around $(83 \pm 5)$\,$M_{\odot}$ and $(82 \pm 5)$\,$M_{\odot}$ for the
binary components \citep{Rauw05}, what certainly qualifies it among
the most massive binary systems in our Galaxy.  No significant flux
variability or orbital periodicity was found in the corresponding data
sets, neither for the Cygnus\,OB2 nor for the Westerlund\,2 regions.
Unless such an orbital period is detected in future datasets, it would
be very difficult or impossible for the current generation of
instruments to distinguish whether the radiation observed from these
associations is coming from a isolated binary system or rather is
generated as a collective effect of the whole cluster. Thus, 
 a direct measurement of single WR binary systems, to
explore whether they are able to produce high-energy $\gamma$-ray
emission in an isolated condition, is worth pursuing. After briefly
motivating the scenario for $\gamma$-ray emission from isolated binary
systems, we present the results of MAGIC observations of two such
systems.

\section{High-energy $\gamma$-rays from WR binaries: candidates, observations and results} 

For the present investigation, and apart from other technical
considerations like a favorable declination, we looked for candidates
with non-thermal emission, indicating the presence of
relativistic electrons, and for which the geometry of the colliding
wind region is established. The two selected systems
--WR\,146 and WR\,147-- have been resolved using the Very Large Array
(VLA) and the Multi-Element Radio Linked Interferometer (MERLIN) in
two sources each and at least for WR\,147, where the most detailed
model is currently available, the system was predicted to be a powerful
MAGIC source for most of the orbital period \citep{Reimer06}.

For the angular resolution of the MAGIC telescope ($\sim0.1^\circ$)
the colliding wind zone will not appear to be spatially resolved,
presenting individual colliding wind binary systems as point-source
candidates at the $\gamma$-ray sky. Thus we have searched for point
sources in the direction of these two binaries.

\subsection{WR\,147}
WR\,147 \citep[see e.g.,][and references therein]{Setia01}, among the
closest and brightest systems that show non-thermal radio emission in
the cm band, is composed of a WN8(h) plus a B0.5\,V star with a
bolometric luminosity of $L_{\rm bol} = 5\times 10^4 L_{\sun}$ and
effective temperature $T_{\rm eff}=28500$\,K, i.e. thermal photon
energy of about $\epsilon_T\approx 6.6$\,eV.  At a distance of 650\,pc
the implied binary separation is estimated to be 417\,AU. The mass
loss rates ($\dot M_{\rm WR}=2.5\times 10^{-5} M_{\sun}$/yr, $\dot
M_{\rm OB}=4\times 10^{-7}M_{\sun}$/yr) and wind velocities ($v_{\rm
WR}=950$\,km/s, $v_{\rm OB}=800$\,km/s) place the stagnation point at
$6.6\times 10^{14}$\,cm. This is in fact in agreement with MERLIN
observations, which show a northern non-thermal component and a
southern thermal one with a separation of ($575\pm 15$)\,mas
\citep{Churchwell92,Williams97}. This radio morphology and spectrum
support a colliding wind scenario \citep{Williams97}, whose collision
region has also been detected by the Chandra X-ray telescope
\citep{Pittard02}.

\begin{figure}[t!]
\epsscale{1.0}
\plotone{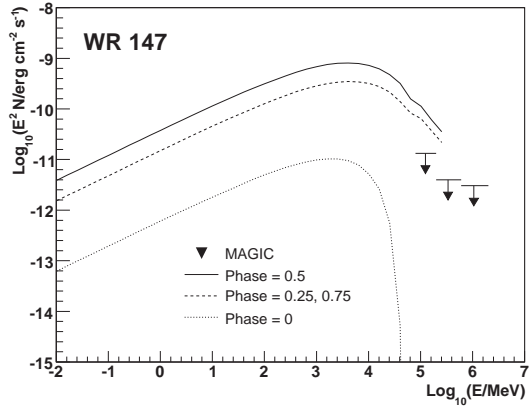}
\caption{Inverse Compton (IC) spectra of WR\,147 for orbital phases 0, 0.25,
0.5 and 0.75, neglecting any eccentricity of the system and assuming
$i=90$\,deg, from \citet{Reimer06}. 
$\gamma\gamma$ pair production absorbs not
more than $\leq$0.3\% ($>50$\,GeV) and $\leq$18\% ($>100$\,GeV) of the
produced flux at orbital phases 0.25 and 0.5, respectively. No
absorption takes place at phase 0. MAGIC upper limits on this system
are marked.  }
\label{fig:wr147}
\end{figure}

The non-thermal flux component can be well fitted by a power law with
spectral index $\alpha=-0.43$.  Neither the eccentricity nor the
inclination of the system are known due to the very long 
orbital period \citep[1350 yr, as derived by][]{Setia01}.
\citet{Reimer06} have provided a detailed modeling of the
high-energy $\gamma$-ray emission expected from WR\,147. The expected
fluxes are shown in Figure~\ref{fig:wr147}, together with MAGIC upper
limits for which we give further details below.

WR\,147 was observed with the MAGIC telescope between 11 August and 10
September 2007, for a total of 30.3 hours of good data (after quality
cuts removing bad weather runs). The zenith angle of the observations
ranged between 10$^\circ$ and 30$^\circ$, being sensitive to
gamma-rays in the energy range between about 80\,GeV and 10\,TeV. The
observations were carried out in the false-source track (wobble) mode
\citep{Fomin1994}, with two directions at $24^\prime$ distance
east-west of the source direction. The analysis of the data was  
performed with the MAGIC standard analysis chain \citep{Alb08c}, which
combines Hillas image parameters by means of a Random Forest algorithm
\citep{Alb08b} for signal/background discrimination and energy
estimation. The training of the algorithm is done by means of
contemporary data from the Crab Nebula observations and Monte Carlo
simulated gamma-ray events. Since February 2007, MAGIC signal
digitization has been upgraded to 2\,GSample/s Flash Analog-to-Digital
Coverters (FADCs), and timing
parameters are used during the data analysis~\citep{Tescaro2007}. This
results in an improvement of the flux sensitivity from 2.5$\%$ to
1.6$\%$ (at a flux peak energy of 280\,GeV) of the Crab Nebula flux in
50 hours of observations.

\begin{deluxetable}{rrrrrr}
\tablewidth{0pt}
\tablecaption{WR\,147 observation results\tablenotemark{a}
\label{tab:wr147}}
\tablehead{
\colhead{Energy} & \colhead{$N_\textrm{excess}$} &
\colhead{S} & \colhead{U.L.} \\ 
\colhead{$[$GeV$]$}&\colhead{[evts]}
&\colhead{[$\sigma$]}&\colhead{[evts (cm$^{-2}$ s$^{-1}$)]}
}
\startdata
$>$80  & -196$\pm$175  & -1.1 & 150 ($1.1 \times 10^{-11}$) \\
$>$200 &  -92$\pm$89   & -1.0 &  84 ($3.1 \times 10^{-12}$) \\
$>$600 &  -20$\pm$24   & -0.8 &  28 ($7.3 \times 10^{-13}$) 
\enddata
\tablenotetext{a}{From left to right: energy range, number of excess
events, statistical significance of the excess \citep{LiMa1983}, and
signal upper limit for the different observation nights. Upper limits
\citep{Rolke2005} are 95$\%$ confidence level (CL) and are quoted in
number of events and (between brackets) in photon flux units assuming a
Crab-like spectrum \citep{Alb08c}.}
\end{deluxetable}

Searches of gamma-rays from WR\,147 have been performed for three
different energy cuts, namely: above 80\,GeV, above 200\,GeV and above
600\,GeV. In all cases the number of signal candidate events found are
compatible with statistical fluctuations of the expected
background. The obtained upper limits are shown in
Table~\ref{tab:wr147} and Figure~\ref{fig:wr147}, and correspond to
1.5$\%$, 1.4$\%$ and 1.7$\%$ of the Crab Nebula flux for the three
considered energy bins, respectively.

\subsection{WR\,146}

WR\,146 is a similar system: a WC6+O8 colliding-wind binary system also
presenting thermal emission from the stellar winds of the two stars,
and bright non-thermal emission from the wind-collision region
\citep[e.g., see][]{Dougherty96,Dougherty00,OConnor05}.  The period is
estimated to be $\sim$ 300\,yr \citep{Dougherty96} and the estimates
of the distance to the system differ from 0.75\,kpc to 1.7\,kpc
\citep[see][and references therein for other system parameters]{Setia01}.

WR\,146 is located $\sim 0.7^\circ$ away from the unidentified VHE
$\gamma$-ray source TeV J2032+4130 and was observed with MAGIC within
the observation program devoted to this source \citep{Alb08a}, albeit
with a reduced sensitivity. The total effective exposure, which
accounts for the loss of sensitivity of off-axis observations and
camera illumination during moonlight observations \citep{Alb07}, is
44.5 hours, obtained between 2005 and 2007. At MAGIC site, WR\,146
culminates at 14$^\circ$ and the observations were carried out at
zenith angles between 14$^\circ$ and 44$^\circ$ \citep[see][for
details]{Alb08a}. The data analysis follows the standard MAGIC
analysis chain. Since most of the data are acquired with 300 MHz FADCs,
image timing parameters are not used in this analysis.

\begin{deluxetable}{rrrrrr}
\tablewidth{0pt}
\tablecaption{WR\,146 observation results\tablenotemark{b}
\label{tab:wr146}}
\tablehead{
\colhead{Energy} & \colhead{$N_\textrm{excess}$} &
\colhead{S} & \colhead{U.L.} \\ 
\colhead{$[$GeV$]$}&\colhead{[evts]}
&\colhead{[$\sigma$]}&\colhead{[evts (cm$^{-2}$ s$^{-1}$)]}
}
\startdata
$>$80   & 264$\pm$97 &  2.7 &  840 ( $3.5 \times 10^{-11}$) \\
$>$200  & 133$\pm$67 &  2.5 &  487 ( $7.7 \times 10^{-12}$) \\
$>$600  & -21$\pm$26 & -0.8 &   46 ( $5.6 \times 10^{-13}$) 
\enddata
\tablenotetext{b}{See explanation in Table~\ref{tab:wr147}}
\end{deluxetable}

The result of the searches of gamma-rays from WR\,146 for three
different energy cuts (above 80\,GeV, above 200\,GeV and $>$600\,GeV)
are shown in Table~\ref{tab:wr146}. As for the case of WR\,147, all
measured signal candidates are consistent with background fluctuations
and the upper limits (corresponding to $5.0\%$, $3.5\%$ and $1.2\%$ of
the Crab Nebula flux) are presented. At the lower energy bins we see a
positive number of excesses at the $\sim 2.5\sigma$ level. However,
the present data are too scarce to establish if this comes from a
gamma-ray signal or background fluctuations. Future observations will
shed light on this issue.

\section{Concluding remarks} 

Our search for VHE $\gamma$-ray emission from two archetypical cases
of WR binaries produced the first bounds on such systems. These bounds
constrain theoretical models (or, assuming correctness of the models,
their internal parameters, such as the -unknown- orbital phases of the
systems). The establishment of WR binaries as VHE $\gamma$-ray sources
is yet pending.

The case for WR\,147 as a potential $\gamma$-ray source for MAGIC was
theoretically established before, as shown in the corresponding curves
of Figure~\ref{fig:wr147} from \citet{Reimer06}, for most of the
orbital phases. The validity of this model is baselined on an assumed
ensemble of orbital parameters, which are still unknown for this
system. For instance, ignorance of its current phase as well as of its
eccentricity and inclination makes a direct ruling out of this model
impossible, although that the presented scenario could nominally
survive only for phases close to 0, defined where the line of sight
encounters first with the WN8 and then the B0.5V star. The MAGIC
observations show that irrespective of phase, GLAST should see a flux
cutoff well within its range of detectability in the tens of GeV
regime, if it is able to detect the stars at all.


{\it We acknowledge A. Reimer for providing us the theoretical curves
depicted in Figure 1. We also would like to thank the Instituto de
Astrofisica de Canarias for the excellent working conditions at the
Observatorio del Roque de los Muchachos in La Palma. The support of
the German BMBF and MPG, the Italian INFN and Spanish CICYT is
gratefully acknowledged. This work was also supported by ETH Research
Grant TH 34/043, by the Polish MNiSzW Grant N N203 390834, and by the
YIP of the Helmholtz Gemeinschaft.}

\end{document}